# Analyzing Stop-and-Go Waves by Experiment and Modeling

A. Portz and A. Seyfried

Jülich Supercomputing Centre, Forschungszentrum Jülich GmbH
52425 Jülich, Germany
Corresponding author: a.portz@fz-juelich.de

**Abstract** The main topic of this paper is the analysis and modeling of stop-and-go waves, observable in experiments of single lane movement with pedestrians. The velocity density relation using measurements on a 'microscopic' scale shows the coexistence of two phases at one density. These data are used to calibrate and verify a spatially continuous model. Several criteria are chosen that a model has to satisfy: firstly we investigated the fundamental diagram (velocity versus density) using different measurement methods. Furthermore the trajectories are compared to the occurrence of stop-and-go waves qualitatively. Finally we checked the distribution of the velocities at fixed density against the experimental one. The adaptive velocity model introduced satisfies these criteria well.

## Introduction

To enhance crowd management or to improve escape route systems computer simulations are used to describe the dynamics of pedestrians [1-4]. Here one main aim is to analyze when and where long lasting jams or congestions occur and how they could be prevented. But tests to show whether the models used for simulation are able to describe e.g. formation of jams or the degree of congestion are still in their infancy and suffer from a poor experimental data base. In particular specifications of the density, where jamming starts, are inconsistent and range from 3.8 to 10 [Pers. per m²] for planar corridors [2]. Moreover there is no consensus in the community as to which criteria have to be chosen for model verification [5]. To improve this situation we present experimental results about stop-and-go waves in single file movement. Even in this simple case pedestrians interact in many ways and not all factors, which have an effect on their dynamics, are known. One proven factor is the cultural differences occurring in the fundamental diagram [6]. Along with a new modeling approach we introduce quantitative criteria to calibrate and test models regarding characteristics of stop-and-go waves. For model-ing pedestrian motion, we start with the simplest case: single lane movement. If the model is able to describe the dynamics of pedestrians quantitatively and qualitatively for that simplified case, it is a good candidate for adaption to two-dimensional situations. In the past we followed different modeling approaches continuous in space and validated them with empirical data [7]. Only one model satisfied our criterion, reproduction of the fundamental diagram. This model, with adaptive velocity, is now discussed in detail regarding stop-and-go waves. At first we systematically analyze the experimental data to get significant criteria for stop-and-go waves. This is done in the first section. After this the adaptive velocity model is introduced and validated. To avoid deviations due to different measurement methods [8] we used the same methods for the comparison of experimental data with model results.

## Experimental Data

We obtained empirical data from two experiments of single lane movement we had performed in previous years. There a corridor with circular guiding was built to realize periodic boundary conditions. The density was varied by increasing the number of the test persons in the corridor. The experiment was performed on two occasions using different corridor lengths and test persons.

The first experiment was conducted at the auditorium Rotunde of the Jülich Supercomputing Centre (JSC), Forschungszentrum Jülich [9]. The group of pedestrians was composed of students and staff of JSC. The length of the measured section was $l = 2$ [m] and the whole corridor was 17.3 [m]. We executed runs with N=15, 20, 25, 30 and 34 pedestrians in the passageway.

The second experiment was performed in 2006 in the wardroom of Bergische Kaserne Düsseldorf [8]. The test group was of soldiers. Here the length of the system was about 26 [m], with a $l = 4$ [m] measurement section. We were able to do several runs with up to 70 pedestrians.

Detailed information about the experimental set-up and analysis, is given in [8, 9]. In the first experiment the measurement of velocity and density was done manually. For the second experiment we used advanced cameras and the tool PeTrack for automatic extraction of pedestrian trajectories from video recording [10].

### *Fundamental Diagram*

Firstly we compare the data of both experiments to test whether system's length and composition of the test group lead to differences in the results. For this, the velocity $v_i$ is calculated using the entrance and exit times $t_i^{in}$ and $t_i^{out}$ of pedestrian $i$ into and out of the measurement section $l$,

$$v_i = \frac{l[m]}{(t_i^{out} - t_i^{in})[s]}.$$

To avoid discrete values of the density leading to a large scatter, we define the density by

$$\rho(t) = \frac{\sum_{i=1}^{N} \Theta_i(t)}{l[m]},$$

where $\Theta_i(t)$ gives the fraction of the space between pedestrian $i$ and $i+1$, which can be found inside the measurement area, see [8]. $\rho_i$ is the mean value of all $\rho(t)$ for $t \in [t_i^{in}, t_i^{out}]$

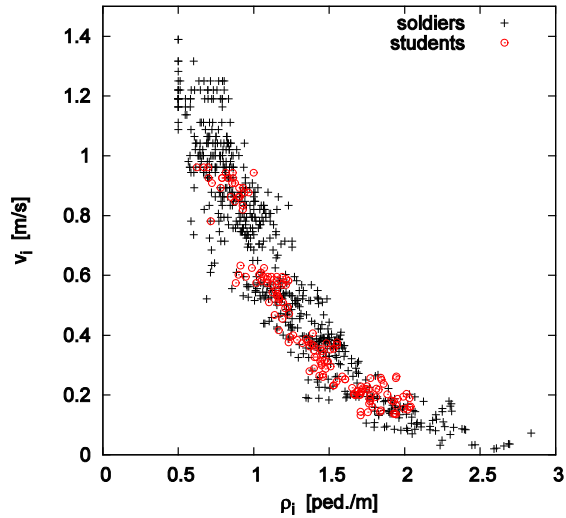

**Fig. 1. Fundamental diagrams for both experiments (soldiers and students) for the single file movement.**

The results are shown in Fig. 1. The fundamental diagram of the students lies on the same curve as the one of the soldiers. This indicates that the data gained by the experiments are representative and reproducible, so we can treat data sets equally for the next step.

## *Stop-and-Go Waves*

In Fig. 2 the *x*-component of trajectories of the pedestrians are plotted against time. For the extraction of the trajectories, the pedestrians' heads were marked in advance and tracked. Backward movement in the trajectories leading to negative velocities is caused by head movement of the pedestrians during a standstill. Irregularities in the trajectories increase with increasing density. In both plots the stop-and-go waves occur throughout the experiments with the wave propagating opposite to the movement direction. Stopping is first observed during the runs with 45 pedestrians, at 70 pedestrians they can hardly move forward. Stop-and-go phases occur simultaneously. Some pedestrians stop, but elsewhere pedestrians could walk slowly.

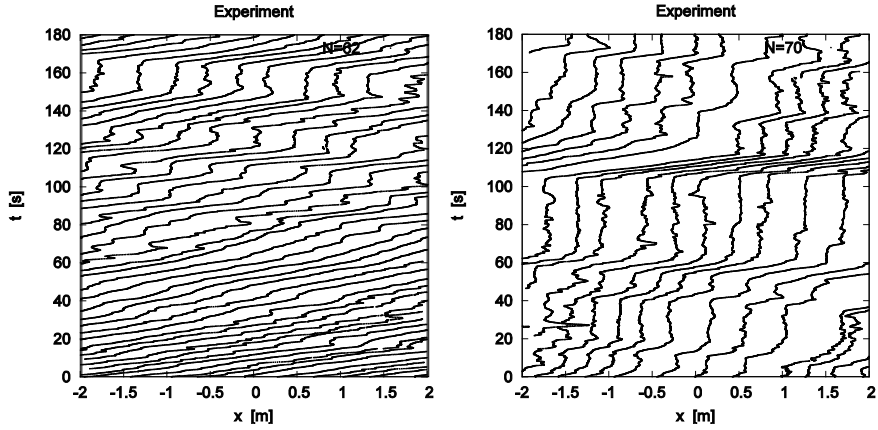

**Fig. 2.** Stop-and-go waves in trajectories of the experiment for 62 and 70 pedestrians.

These two phases, separated in space and time, are also observable at a fixed density. Before we discuss this in detail, we calculate the fundamental diagram with a different method of measurement. We wish to determine the fundamental diagram on the scale of individuals. To achieve this we apply the Voronoi density method [11] to the one dimensional case. In one dimension a Voronoi cell is defined by the center $z_i$ between pedestrian position $x_i$ and $x_{i+1}$

$$z_i = \frac{x_{i+1} + x_i}{2} \quad , \quad L_i = z_{i+1} - z_i \quad \text{and} \quad \rho_i(x) = \begin{cases} \frac{1}{L_i}, & x \in [z_i, z_{i+1}[ \\ 0, & \text{otherwise} \end{cases}.$$

The fundamental diagram using the Voronoi method is shown in the left part of Fig. 3. Regular stops occur at densities higher than 1.5 pedestrians per meter. On the right side of Fig. 3 the distribution of the velocities for fixed densities from 1.8 to 2.6 pedestrians per meter are shown. There is a continuous change from a single peak near $v = 0.15$, to two peaks, to a single peak near $v = 0$. The right peak represents the walking pedestrians, whereas the left peak represents the stopping pedestrians. For a density between 2.0 and 2.2 pedestrians per meter both peaks coexist. Several causes could be responsible for the double peak structure of the velocity distribution. One is, that different pedestrians react to the situation differently and have different personal space requirements. Some pedestrians prefer a larger personal space than other. Also differences in the reaction time could be re-sponsible. These differences in the perception and personal space requirements could be combined with differing step phases and stopping stances.

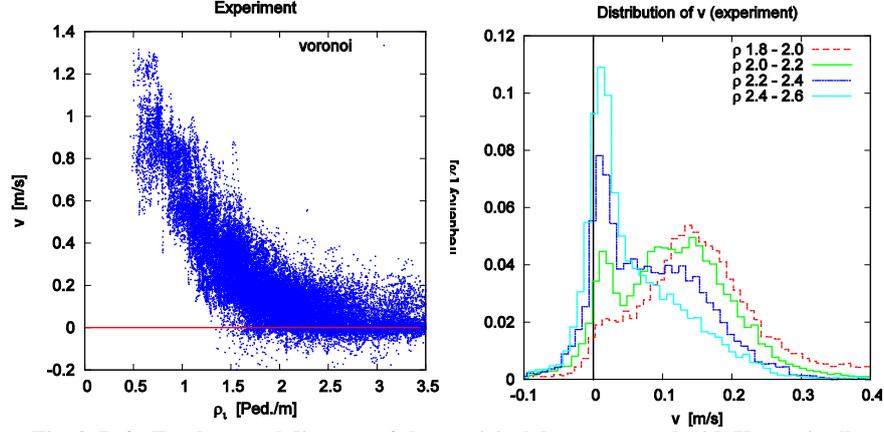

**Fig. 3. Left: Fundamental diagram of the empirical data measured with Voronoi cells. Right: Distribution of *v* at fixed densities.**

## Modeling

### *Adaptive Velocity Model*

In this section we introduce the adaptive velocity model, which is based on an event driven approach. A pedestrian can be in different states and a change between these states is called event. The calculation of the velocity of each pedestrian is straightforward and depends on these states. The model was derived from force based models, where the dynamics of pedestrians are given by the following system of coupled differential equations

$$m_i \frac{dv_i}{dt} = F_i \quad \text{with} \quad F_i = F_i^{drv} + F_i^{rep} \quad \text{and} \quad \frac{dx_i}{dt} = v_i,$$

where $F_i$ is the force acting on pedestrian *i*. The mass is denoted by $m_i$, the velocity by $v_i$ and the current position by $x_i$. $F_i$ is split into a repulsive force $F_i^{rep}$ and a driving force $F_i^{drv}$. The dynamic is regulated by the interrelation between driving and repulsive forces. In our approach the role of repulsive forces are replaced by events. The driving force is defined as

$$F_i^{drv} = \frac{v_i^0 - v_i}{\tau_i},$$

where $v_i^0$ is the desired velocity of a pedestrian and $\tau$ their reaction time. By solving the differential equation

$$\frac{dv_i}{dt} = F_i^{drv} \quad \Rightarrow \quad v_i(t) = v_i^0 + \exp\left(-\frac{t}{\tau_i}\right),$$

the velocity function is obtained. This is shown in Fig. 4 together with the parameters governing the pedestrians' movement.

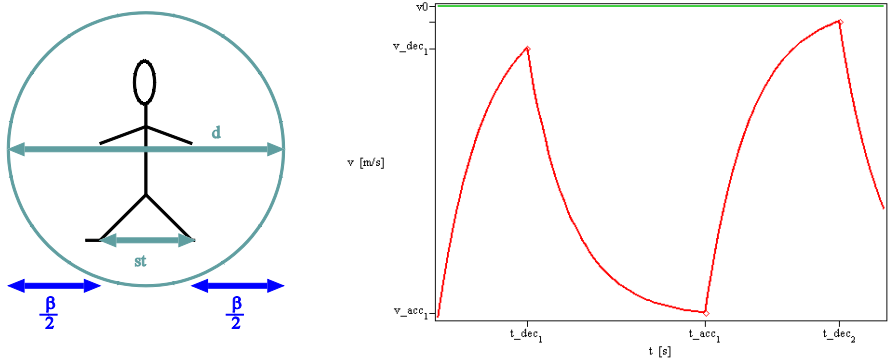

**Fig. 4.** Left: connection between the required space *d*, the step length *st* and the safety distance *β*. Right: the adaptive velocity with acceleration until $t_{dec1}$ than deceleration until $t_{acc1}$, again acceleration until $t_{dec2}$ and so on.

In this model pedestrians are treated as hard bodies with a diameter $d_i$ [12]. The diameter depends linearly on the current velocity and is equal to the step length *st* in addition to the safety distance *β*

$$d_i(t) = e + f\, v_i(t) = st_i(t) + \beta_i(t). \tag{1}$$

Based on [13] the step length is a linear function of the current velocity with following parameters:

$$st_i(t) = 0.235[m] + 0.302[s]v_i(t). \tag{2}$$

*e* and *f* can be specified through empirical data. Here *e* is the required space for a stationary pedestrian and *f* describes the additional space a moving pedestrian requires. For *e* = 0.36 [m] and *f* = 1.06 [s] the last equations (1) and (2) can be summarized to

$$\beta_i(t) = d_i(t) - st_i(t) = 0.125[m] + 0.758[s]v_i(t).$$

A pedestrian accelerates to his/her desired velocity $v_i^0$ until the distance to the pedestrian in front ($\Delta x_{i,i+1}$) is smaller than the safety distance. From this time on, he/she decelerates until the distance is larger than the safety distance and so on. Via $\Delta x_{i,i+1}$, $d_i$ and $\beta_i$ events can be defined: deceleration (3) and acceleration (4). To implicitly include a reaction time and to ensure good computational performance for high densities, no events are explicitly calculated. Instead in each time step of $\Delta t = 0.05$ seconds, it is checked whether an event has taken place and $t_{dec}$, $t_{acc}$ or $t_{coll}$ are set to *t* accordingly. The discreteness of the time step could lead to configurations where overlapping occurs. To guarantee a minimal volume exclusion, case (5) is included, in which the pedestrians are too close to each other and have to stop. To avoid artifacts related to the update procedure a recursive update procedure is necessary: Each person is advanced one time step according to equa-tions (3-5). If after this step a pedestrian is in a different state because of the new distance to the pedestrian in front, the velocity is set according to this state. Then the state of the next following person is reexamined. If the state is still valid the update is completed. Otherwise, the velocity is again calculated and so on.

$$t = t_{dec}, \quad \text{if:} \quad \Delta x_{i,i+1} - 0.5*(d_i(t) + d_{i+1}(t)) \leq 0 \qquad (3)$$

$$t = t_{acc}, \quad \text{if:} \quad \Delta x_{i,i+1} - 0.5*(d_i(t) + d_{i+1}(t)) > 0 \qquad (4)$$

$$t = t_{coll}, \quad \text{if:} \quad \Delta x_{i,i+1} - 0.5*(d_i(t) + d_{i+1}(t)) \leq -\beta_i(t)/2 \qquad (5)$$

To incorporate the dispersion in the characteristics and behavior of pedestrians we choose the following parameters from a normal distribution, $\tau_i \sim N(1.0, 0.1)$, $e_i \sim N(0.36, 0.1)$ and $f_i \sim N(1.06, 0.5)$.

## *Fundamental Diagram*

The fundamental diagram of our modeled and empirical data is displayed in Fig. 5. The adaptive velocity model yields the right relation between velocity and density.

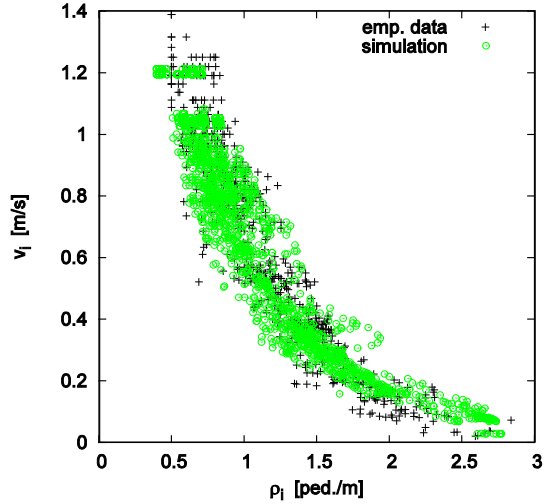

**Fig. 5. Validation of the modeled fundamental diagram with the empirical data.**

We also tested different time steps $\Delta t$ and $\tau_i = 1.0$, $e_i = 0.36$, $f_i = 1.06$ for all pedestrians. For $\Delta t$ from 0.001 to 0.1[s] the included reaction time has no significant influence on the shape of the fundamental diagram. To avoid interpenetrations of pedestrians we choose $\Delta t = 0.05$[s] for further simulations. Variation of the personal parameters $\tau_i$, $e_i$ and $f_i$ affects the scatter of the fundamental diagram.

*Stop-and-Go Waves*

For a qualitative comparison we plot the modeled trajectories in Fig. 6. Stop-and go waves occur as in the experiment, see Fig. 2. The pattern as well as the change of the pattern from $N = 62$ to $N = 70$ are in good agreement with the experimental results. Even the start and stop phases match qualitatively. However the phases of stop and go traffic appear more regular in the modeled trajectories.

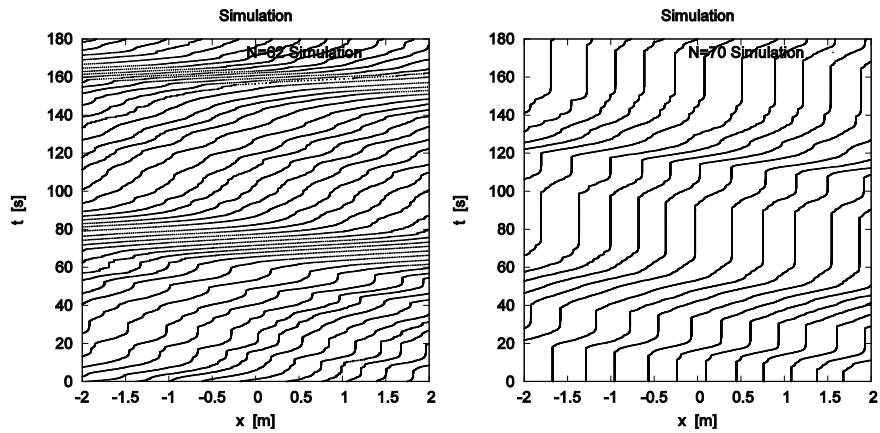

**Fig. 6. Modeled trajectories for 62 and 70 pedestrians.**

Furthermore we examine the double peak structure, exhibited by the experimental data, see Fig. 3. This acts as a quantitative criterion.

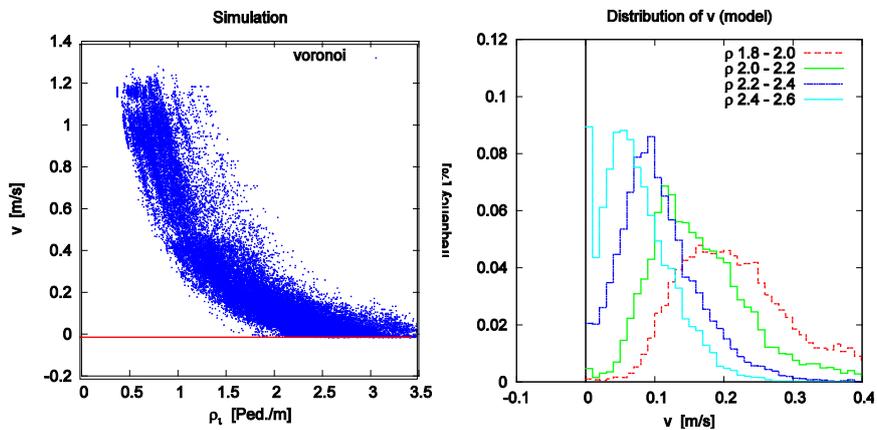

**Fig. 7. Left: Fundamental diagram of the modeled data measured with Voronoi cells. Right: Distribution of $v$ at fixed densities.**

There is also a double peak in the distribution of the modeled velocities, see Fig. 7. The velocity distribution is similar but not identical to the empirical one. The double peak structure given by the height and position of the peaks differ from experimental results. These differences could be caused by the irregularities of the experimental data. This is also recognizable in the comparison of the trajectories in Fig. 6 and Fig. 2. To avoid these irregularities a step-detection and a cleaning of the empirical trajectories is necessary. We will consider this in future work.

## Conclusion and Perspectives

Choosing an adequate distribution of individual parameters, the adaptive velocity model is able to generate the fundamental diagram and to create stop-and-go waves without unrealistic phenomena, like overlapping or interpenetrating pedestrians. Even the change of the characteristics of stop and go waves for varying densities is in good agreement with the experiments. For a more detailed analysis of stop-and-go waves we need to clean the experimental data, so that the irregularities caused by the self dynamic of pedestrians' head vanish. In the future we will try to get a deeper insight into to occurrence of stop-and-go waves with other criteria like average stopping time. Furthermore we plan to include steering of pedestrians. For the validation of these models more test cases, like flow characteristics at bottlenecks and junctions will be used..

**Acknowledgement:** This study was supported by the German Government's high-tech strategy, the Federal Ministry of Education and Research (BMBF). Program on "Research for Civil Security - Protecting and Saving Human Life". Execution of experiments was supported by the German Research Foundation (DFG) KL 1873/1-1 and SE 1789/1-1.